# Probing shells against buckling: a non-destructive technique for laboratory testing

**J. Michael T. Thompson**
*Department of Applied Maths and Theoretical Physics,*
*University of Cambridge, Cambridge CB3 0WA, UK*

**John W. Hutchinson**
*School of Engineering and Applied Sciences,*
*Harvard University, Cambridge, MA 02138*

**Jan Sieber**
*CEMPS, University of Exeter,*
*Exeter EX4 4QF, UK*

**Abstract.** This paper addresses testing of compressed structures, such as shells, that exhibit catastrophic buckling and notorious imperfection sensitivity. The central concept is the probing of a loaded structural specimen by a controlled lateral displacement to gain quantitative insight into its buckling behaviour and to measure the energy barrier against buckling. This can provide design information about a structure's stiffness and robustness against buckling in terms of energy and force landscapes. Developments in this area are relatively new but have proceeded rapidly with encouraging progress.

Recent experimental tests on uniformly compressed spherical shells, and axially loaded cylinders, show excellent agreement with theoretical solutions. The probing technique could be a valuable experimental procedure for testing prototype structures, but before it can be used a range of potential problems must be examined and solved. The probing response is highly nonlinear and a variety of complications can occur. Here, we make a careful assessment of unexpected limit points and bifurcations, that could accompany probing, causing complications and possibly even collapse of a test specimen. First, a limit point in the probe displacement (associated with a cusp instability and fold) can result in dynamic buckling as probing progresses, as demonstrated in the buckling of a spherical shell under volume control. Second, various types of bifurcations which can occur on the probing path which result in the probing response becoming unstable are also discussed. To overcome these problems, we outline the extra controls over the entire structure that may be needed to stabilize the response.

## 1. Introduction

Many general discussions of shell buckling start with a sketch like the one below in figure 1(a), showing the potential energy, $W$, of the structure and its loading device as a function of a lateral deflection, $D$, and the magnitude of a controlled loading parameter, $G$. This parameter might be a generalised force applying 'dead' loading or its generalised displacement applying 'rigid' loading. The typical example of a qualitative diagram in figure 1(a) is taken from a 1960's thesis [Thompson, 1961].

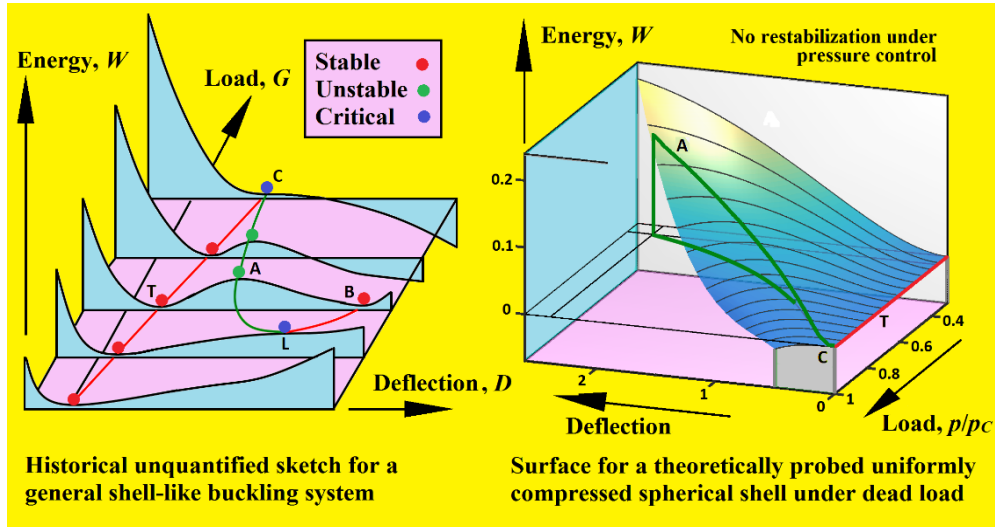

Fig 1. Two energy landscapes (a) A historical unquantified sketch for a general post-buckling system (b) An accurate calculated surface for a complete spherical shell under dead uniform pressure loading using a theoretical simulation of a rigid point probe.

Energy levels at a given $G$ are measured from the datum of the trivial unbuckled solution drawn in red. The heights of the green balls display the energy levels of the unstable post-buckling path (also in green) falling with decreasing $G$ from the bifurcation point. These are the energy barriers which must be surmounted by any disturbances that might trigger an escape from the potential well of the trivial solution. Often the falling post-buckling path will eventually curve upwards at a fold (limit point, L in figure 1) as happens with the spherical shell under controlled external volume. This is shown here, with the red ball at B displaying the energy of the large-deflection stable post-buckling state. All the ball heights have been implicitly evaluated in many studies of the post-buckling, with varying degrees of accuracy, and they do have a precise well-defined meaning: namely the energy of the post-buckling equilibrium states. But the curves between them are totally undefined, because the continuous (or finely discretized) shell has an infinite (or large) number of degrees of freedom. So, between balls there exist an infinite number of possible curves depending on the path chosen in this multi-dimensional state space.

Recently, emphasis has been given both theoretically and experimentally to the response of compressed shells to a lateral point load, called a probe or poker. This probe, invariably driven by a rigid displacement-controlled device, is intuitively placed to directly induce what, under a dead probe, would be a snap-buckle towards large post-buckling deflections. This probing pushes the shell along a path through its state space towards and usually hitting the green unstable solution: and sometimes (but not always) continuing to the red stable solution as well, though this is less important. Intermediate points on the path are not of course equilibrium points of the original un-probed shell, but they and their energies are easily obtained (by this probing) on what would seem to be an ideal *quantified* path for display purposes. Indeed, these probed paths can be expected to (initially) follow some sort of energy valley through the state space. In comparison with the unquantified sketch of Figure 1(a), we present on figure 1(b) a quantified theoretical result, using a point probe, for a uniformly compressed spherical shell [Hutchinson & Thompson, 2017a,b] which we shall be discussing later.

## 2. The Basic Probing Method

The basic method as proposed by Thompson [2015] and Thompson & Sieber [2016] can be summarized as follows. We first note that the buckle shape at the lowest energy barrier determined by Horak et al [2006] for an axially loaded cylindrical shell looks remarkably like the small dimple that might be pressed into the cylinder by a researcher's finger. This immediately suggested a new form of experimental test on a compressed shell (of any shape) in which a lateral point load is applied by a rigid loading device. This would seem to be a useful type of non-destructive and non-invasive test for a shell to determine its shock-sensitivity. The type of test configuration that we had in mind is illustrated, for a cylindrical shell, in figure 2 where the lateral 'probe' moves slowly forward along a fixed line driven by a screw mechanism.

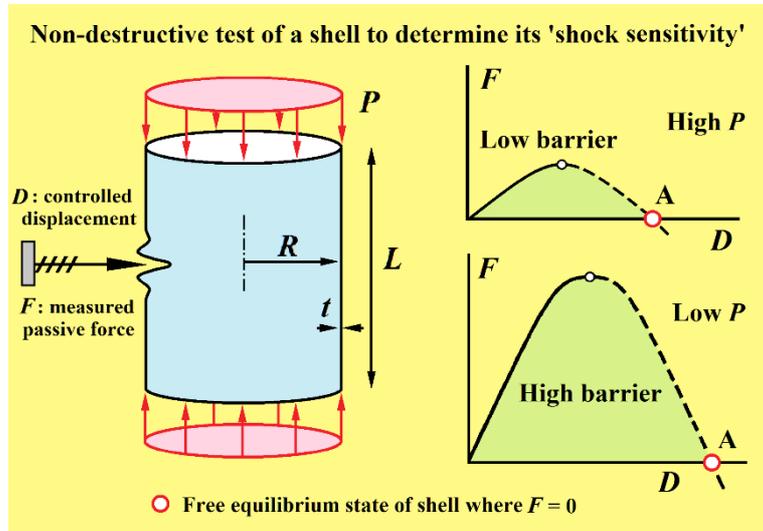

Fig 2 An impression of the proposed experimental procedure in which a rigid probe is used to construct a lateral-load versus lateral-displacement graph, $F(D)$. This graph ends with $F = 0$ at a free equilibrium state of the shell, and the area under the curve gives the corresponding energy barrier. Note that at $F = 0$ the rigid probe is stabilizing a state that would otherwise be unstable for the free shell.

Here we have the controlled displacement, $D$, producing a passive reactive force from the shell, $F$, which is sensed by the device, giving finally the plot of $F(D)$. This is all to be repeated at prescribed values of the axial compressive load, $P$, which might itself be applied in either a dead or rigid manner. In the simplest scenario, the $F(D)$ graphs might be expected to look like those sketched on the right-hand side, the top for a high value of $P$, the bottom for a low value of $P$ (but higher than any minimum of the unstable post-buckling path).

When the test reaches the point A, where $F = 0$, we have located a free equilibrium state of the shell, hopefully, the desired lowest mountain pass. If we finally evaluate the area under the $F(D)$ curve, this will give us the energy barrier that has to be overcome to cause the shell to collapse at the prescribed value of $P$.

It is interesting to note, here, that Takei *et al* [2014] use an imposed lateral displacement in their computations to find the unstable Maxwell state of a thin film: this Maxwell load has proved of great relevance in a wide range of problems [Hunt, *et al*, 2003; Thompson & van der Heijden 2014].

Notice that if the curves have the forms drawn (with no folds or bifurcations), we can be sure that the shell will remain stable up to A under the controlled $D$. If the probe controls $D$ in both directions, rigidly holding the shell at the probe point by gluing (welded or fastened), we will be able to pass point A, with the probe then carrying a negative $F$: if however, the probe is just resting against the shell, a dynamic jump from A will be observed, probably damaging the shell due to large bending strains. Clearly full bilateral control is preferred, to prevent this jump, but other factors must be considered. On the negative side, fixing the probe to the shell may itself cause damage, and may restrict the free deformation that we are seeking to observe. In particular, it might also restrict the shell by preventing a rotational instability. Perhaps the ideal solution would be to have an 'equal and opposite' second probe inside the shell at the same point, moving at the same rate as the probe outside.

The various ways that this simple procedure can fail, by for example reaching a vertical tangency or a bifurcation on the $F(D)$ curve before reaching A, are explored in detail by Thompson & Sieber [2016] focusing on simple models of both cylindrical and spherical shells. In particular, they explore the stabilization of the process by means of a secondary probe that is tuned to provide zero force as illustrated in figure 3. We shall be looking more closely at these matters in a later section.

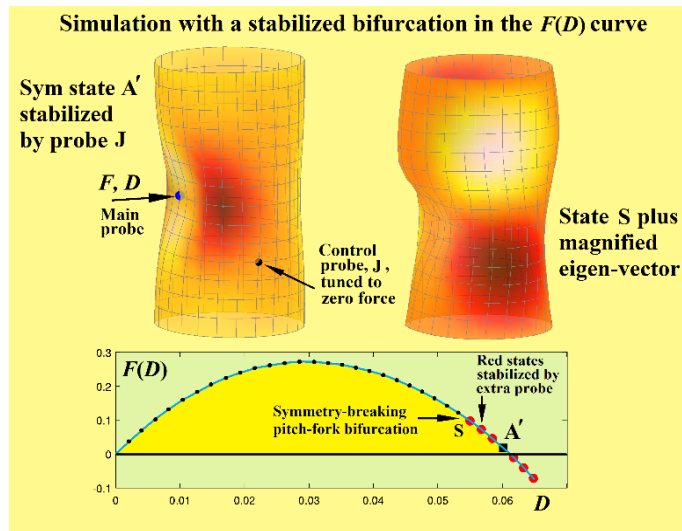

Fig 3 A symmetry-breaking bifurcation found by dynamic shell simulations (Thompson & Sieber [2016]). The illustration of the eigenvector demonstrates how the lateral $F(D)$ curve can be located beyond the bifurcation point, S, by the introduction of a second rigidly controlled point probe, tuned to provide zero force.

## 3. Theoretical results for a spherical shell

Consider, now, the theoretical studies made by Hutchinson & Thompson [2017a, b] using an accurate shell theory formulation (small strain-moderate rotation theory) for the uniformly compressed complete spherical shell developed by Hutchinson [2016]. Probing characteristics from these papers, under controlled (dead) external pressure, are shown in figure 4(a) while the characteristics under controlled (rigid) volume are displayed in figure 4(b). Here the pressure on the shell is written as $p$, with its classical critical value as $p_C$. Likewise, the volume change is $\Delta V$ and $\Delta V_C$.

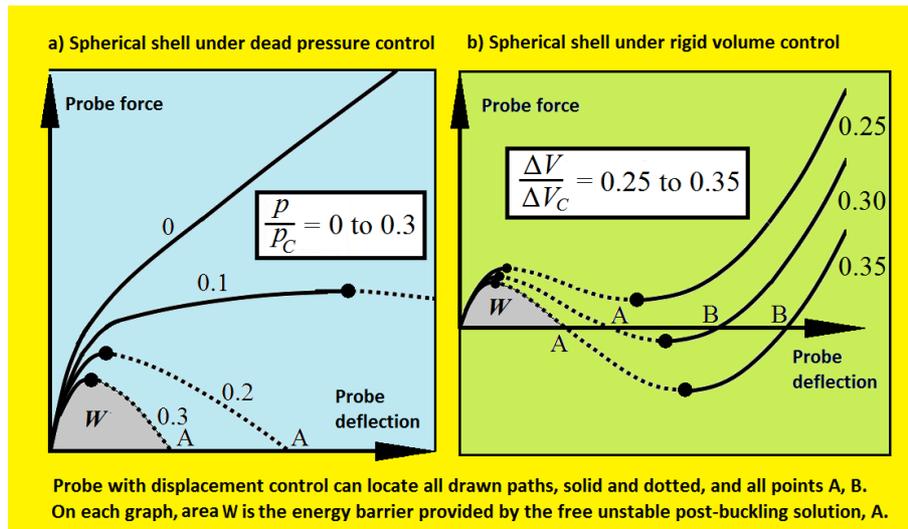

Fig 4. Comparisons between the probing of dead pressure-controlled spheres and rigid volume-controlled spheres. For all the displayed theoretical results a rigid probe could locate all the curves including the free post-buckling states, A and B, and the areas displaying the energy barriers (two of which are coloured in grey).

The significant difference between these two cases arises from the fact that under dead pressure the falling unstable post-buckling solution never restabilizes at a 'lower-pressure buckling load' [Karman & Tsien, 1939, 1941]. On the other hand, the post-buckling path does restabilize at a 'lower-volume buckling' load due to its backwards sweep on the conventional pressure (vertical) versus volume (horizontal) diagram. So, as illustrated in figure 4(b), the probe force turns negative at displacement A but then rises for increasing deflection, crossing zero again at B. With the usual displacement control of a probe attached (or glued) to the shell, all the *drawn* paths would be stable, allowing the location of both the unstable and stable regimes of the free post-buckling solution of the un-probed shell.

The emphasis in the last sentence was on the word *drawn*, and our more recent studies (presented here for the first time) show that the probed shell may not be stable for all D up to state B. Figures 5 and 6 illustrate that a probing experiment will encounter obstacles at larger fixed volumes for a shell with a radius to thickness ratio of R/t = 100 and Poisson's ratio of v = 0.3.

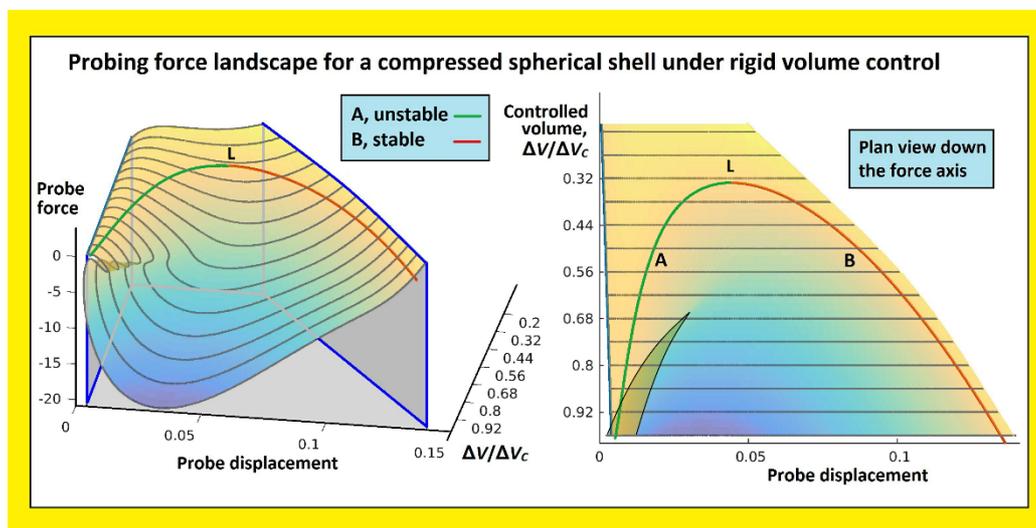

Fig 5. New results for the probing of a complete spherical shell loaded by uniform external pressure. Notice that the unstable post-buckling path, A, is stabilized at point L which is the 'lower buckling point' of this volume-controlled system: and observe the additional vertical tangents of the surface near the critical point (zero displacement, $\Delta V \simeq \Delta V_C$).

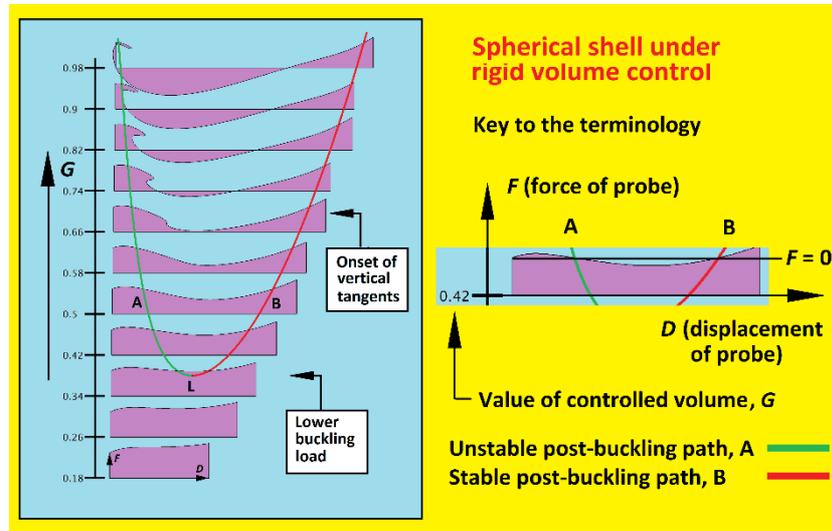

Fig 6. Identifying the features of the spherical shell probing under rigid volume control. Vertical tangents in the force-deflection curves would prevent a simple progression to the large amplitude stable state, B.

Figure 6 shows theoretical probing plots of force, $F$, versus displacement, $D$, for the spherical shell at equal steps (0.18– 0.98) of the controlled volume parameter, $G = \Delta V/\Delta V_C$. All $F(D)$ plots start at $F$ equal to zero, and return to zero whenever they hit the unstable (green) or stable (red) free post-buckling states of the compressed but un-probed shell. The bases of the coloured slabs are drawn, arbitrarily for visual purposes, at $F = -10$. The vertical tangents of the $F(D)$ curves starting at $G \simeq 0.66$, lie beyond the unstable points, A, but they do prevent any simple progression to the stable states, B. We shall discuss this more fully in a later section.

The force landscape in Figures 5 and 6 has been computed as the solution of two coupled third-order nonlinear boundary-value problems for the normalized deflection $d$ and the rotation $\varphi$. Both variables are functions of the latitude angle $\theta$ on the sphere in the interval $[0, \pi/2]$. The precise differential equations are described by Hutchinson [2016] as the result of the assumption of small strain and moderate rotation, combined with the symmetry assumption that the shell stays rotationally symmetric around the pole axis and reflection symmetric with respect to the equator. To enable continuation of the solution through vertical tangents, the pole deflection $D = d(\pi/2)$, which is a parameter of the nonlinear problem entering in the boundary condition, was left as an additional free variable in the nonlinear solver such that the boundary-value problem could be embedded into a curve-tracking algorithm [Dankowicz & Schilder 2013].

In a submitted paper Hutchinson & Thompson [2018] compare the aforementioned results of Horak *et al* [2006] for the energy barrier of an axially loaded cylinder with their own results for the pressurised sphere, obtaining excellent agreement between these two disparate shell structures as shown in figure 7. Similar work on the cylinder by Kreilos & Schneider [2017] should be noted, but their paper does not include results for the barrier.

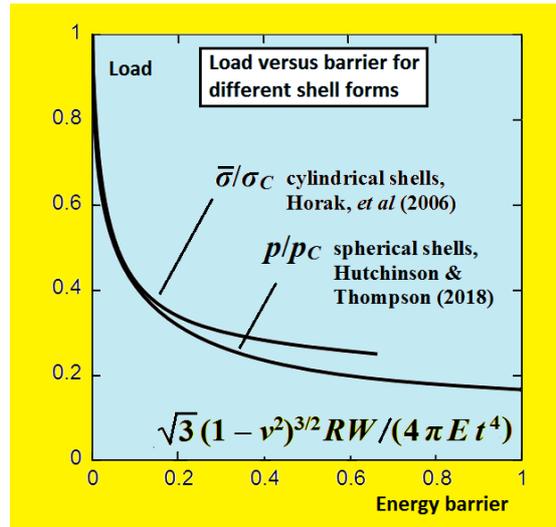

Fig 7 Good agreement between two theoretical load-barrier curves for disparate shell structures. The curve for an axially compressed cylinder is due to Horak et al [2006], while the curve for a pressurized sphere is due to Hutchinson & Thompson [2018]. The dimensionless form of these curves applies to thin cylindrical and spherical shells with radius to thickness ratios greater than about 50.

## 4. Experiments on spherical shells

An experimental study probing hemispherical shells has been conducted by Marthelot *et al* [2017]. Historically it is well established that the cylindrical shell under axial compression and the spherical shell under external pressure have comparably catastrophic and imperfection-sensitive buckling behaviour. The shells tested by Marthelot *et al* were clamped at the equator, then subject to fixed prescribed external pressure (pressure control) achieved using a large reservoir with air as the pressurizing medium, and finally probed at the pole. The shells were made of an elastomer with a typical radius to thickness on the order of 100. Near-perfect shells were manufactured which, when tested under pressure alone, buckled at pressures as high as 80% of the buckling pressure for the perfect shell, $p_C$. Significantly, Marthelot *et al* also manufactured shells with precisely formed geometric dimple imperfections at the pole. The buckling pressure of these shells tested under pressure alone ranged from roughly 60% down to 20% of $p_C$, depending on the imperfection amplitude, and in excellent agreement with numerical buckling calculations accounting for the imperfections. The reversible elasticity of the elastomer allowed each of these shells to be tested at many pressure levels even though the shells collapsed under the prescribed pressure when they buckled.

For the pressurized hemi-sphere subject to the displacement-controlled probing, a dimple buckle forms under the probe confined to the vicinity of the pole prior to collapse. Curves of experimentally measured probe force versus probe displacement for 11 values of prescribed pressure from Marthelot *et al* [2017] are shown in Fig. 7 for a shell with $R/t = 119$ which buckled under pressure alone at $p/p_C = 0.74$.

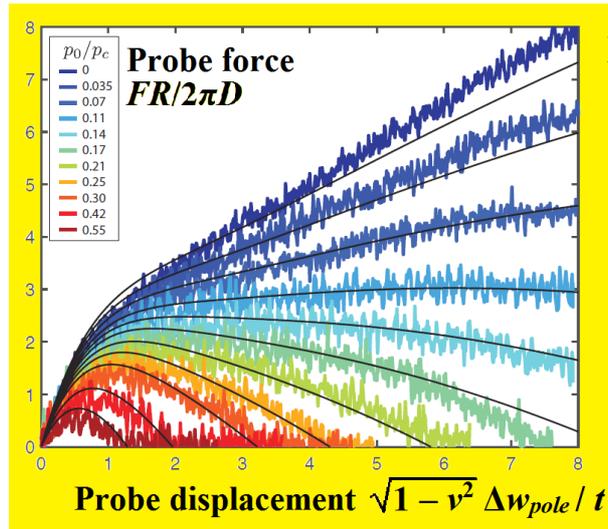

Fig 8 Dimensionless probe force versus dimensionless probe displacement, for 11 values of prescribed pressure from Marthelot *et al* [2017] for a shell with $R/t = 119$. Each experimental curve is the average of 6 probing histories with identical conditions. The solid lines are results computed using the numerical method presented by Hutchinson and Thompson [2017].

Marthelot *et al* [2017] also evaluated the work, *W*, done by the probe to reach the unstable equilibrium state, which for a prescribed pressured is proportional to the area under the curve between the origin and intersection of the probing force with the horizontal axis in the figure. This work is the energy barrier against buckling for a shell loaded to that pressure. A plot of the experimentally measured maximum probe force and the work to reach state A for the tested shell in figure 8 as a function of the prescribed pressure are shown in figures 9(a,b). Included in the figures as solid curves are the theoretical predictions for maximum probe force and the energy barrier. Recent related work on energy barriers of spherical shells is presented by Evkin & Lykhachova [2017].

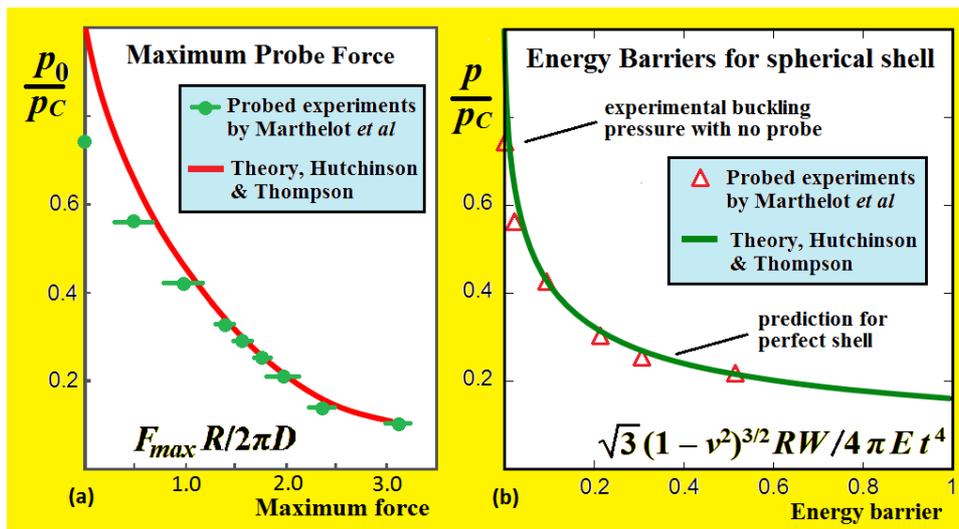

Fig 9 The dimensionless maximum probing force in a) and the dimensionless energy barrier in b) as measured experimentally by Marthelot *et al* [2017] at various values of the prescribed pressure. The theoretical predictions for a perfect shell from Hutchinson & Thompson [2017 & 2018] are shown as solid curves. The shell, the same as that in Fig. 8, buckled at $0.74 p_C$ under pressure alone and was probed at its pole.

## 5. Experiments on cylindrical shells

Recently, Virot *et al* [2017] have made a pioneering series of probing (poking) tests on axially compressed Coke cans. The axial compression is usually rigid (displacement controlled) but they do make a few dead (force controlled) tests to confirm that this makes no essential difference to the measured buckling load. The force control will of course result in the complete destruction of the specimen when it buckles. Meanwhile, the probes are always displacement controlled. The poker tips are steel 'marbles', the diameters of which are shown to be largely irrelevant. Many of the response curves are shown to be reversible, confirming elastic behaviour throughout. Interestingly, at high axial loads several dimples are formed during the dynamic buckling process.

   The authors usually fix the axial loading, and then apply the rigid (displacement controlled) probing, but they show that reversing this process gives the same final result: as in theory, the determined landscape is independent of the control route. The controlled probe displacements do of course first drop to zero at the natural, free-but-unstable post-buckling solution of the shell (A in previous figures), where they typically have a magnitude several times the shell thickness. Tests on a single specimen were used to generate the smoothest landscape, illustrated in figure 10.

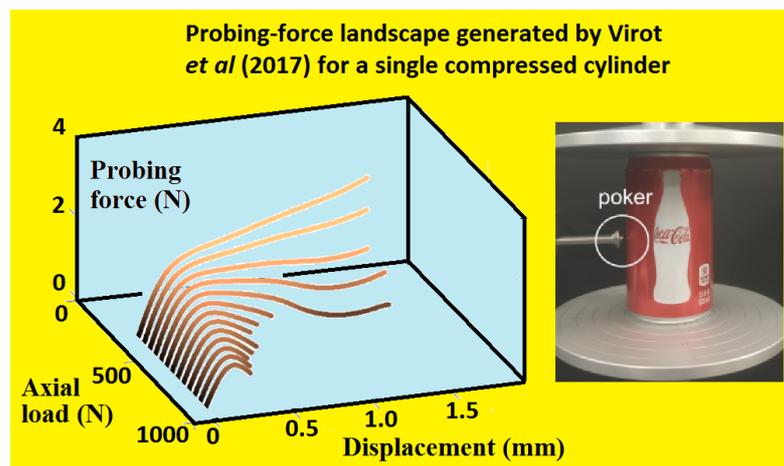

Fig 10. The smoothest experimental force landscape created using a single specimen by Virot et al [2017]. On the left is a photograph of their experimental testing of an axially compressed Coke can.

They do suggest that ridge-tracking along the path of maximum probe force might be useful.

   The protocols of the tests are nicely summarized in their Table 1, which we elaborate below.

| Fixed | Varied | Datasets | Comments |
| --- | --- | --- | --- |
| Axial shortening | Probe deflection | 89 | 'Normal procedure' |
| Axial shortening | Probe deflection | 5 | All on a single specimen |
| Axial Shortening | Probe deflection | 14 | Larger tip diameter |
| Axial load | Probe deflection | 39 | Destruction of specimen |
| Probe deflection | Axial shortening | 17 | Different control route |

Extending figure 4 of their paper into regimes where the ultimate rising stable post-buckling path, B, (under controlled axial shortening) can be determined, the authors produced what they called a stability landscape in the space of the probe force, the probe displacement and the axial shortening. We examine these landscapes further in the following section.

## 6. More about the force landscape

Stability landscapes are conventionally drawn with the two control parameters in the base plane, say the main compression (generalised force or deflection) written as $G$ and the probe deflection, $D$.

From the base plane can be erected either the probe force, or by simple integration, the energy needed to create the probe displacement. The latter has often been drawn or sketched as we illustrated in figure 1, and conveniently displays the energy barrier against final collapse. But we here examine the probe force diagram, as drawn by Virot *et al* [2017], which has other interesting features.

The simplified schematic sketch of figure 11 draws on this experimental work of Virot *et al*, and also on some further test results on a Japanese Asahi can, kindly supplied by the same experimental group.

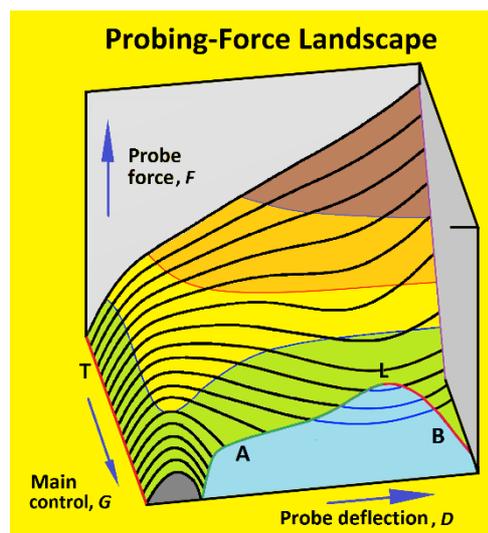

Fig 11. Schematic force landscape based on an Asahi can test. The passive probing force, $F$, is drawn as a surface over the base plane defined by the two control parameters, $G$ and $D$.

Three equilibrium paths of the free un-probed shell, with $F = 0$ are identified on this diagram. The stable (red) trivial path, T, lies along the $G$ axis, the unstable (green) regime of the post-buckling path originating at the (unseen) classical critical point, A, and the re-stabilized (red) regime, B, lying beyond fold L at the lower buckling load. Notice that the notional sketched contours displayed under the surface of the blue 'lake' could only be determined if the probe could exert force in both directions on the shell. However, all the dry shore-line near the lake could be explored without crossing the lake if the experimental controls allowed the following of a curved (or piece-wise linear) path in the $(G, D)$ control plane.

These probe graphs do draw attention to three different tolerances that might be demanded of a practical shell. Under different operating conditions a shell might need a high energy barrier, *W*, a high resistance to lateral forces (the maximum value of *F*), or a large distance to the barrier (the value of *D* at the unstable state A)

## 7. Obstacles to a probing sequence

Assuming that a probe has succeeded in reaching a free equilibrium post-buckling state of a shell-like elastic structure, it is useful to enquire how the path leading from the trivial solution to this state by changing *D* will change qualitatively as the main loading parameter, G, is varied. Of particular interest is whether the path remains 'simple', that is, without any vertical tangents or other events that correspond to buckling of the probed system. To do this, we consider here all of the typical bifurcation events that can generically arise during the probing procedure as we vary one primary parameter (here *D*) to obtain a path and one secondary parameter (here *G*, the main load parameter) to obtain a family of paths.

For a hypothetical elastic structure, figure 12 displays a series of probing tests being made at increasing values of the main load parameter which might be either a dead generalised force or its corresponding deflection. We assume throughout that the probe is acting as a rigid device, with its displacement controlled and is moreover glued to the structure so that it can provide negative force if needed.

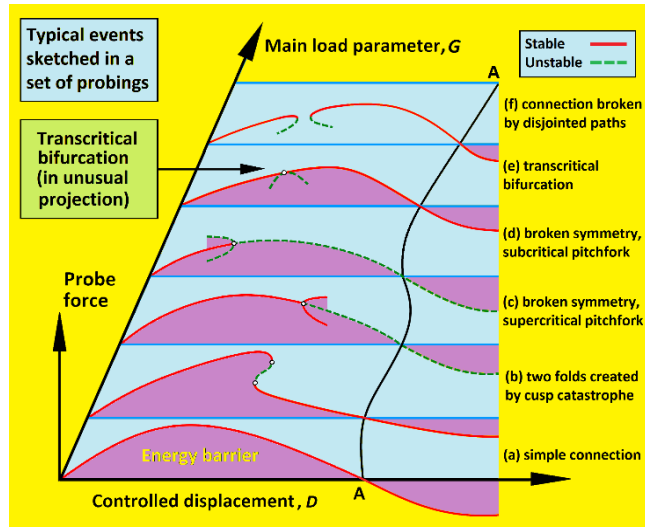

Fig 12 Illustration of all of the generic bifurcations that can arise under the two controls, *G* and *D*, and so disrupt a probing sequence. The appearance of a trans-critical bifurcation in an unfamiliar projection is clarified in figure 13.

In the first (front) probe characteristic there is a direct, simple connection to the unstable post-buckling state, A, and beyond. We focus here on the behaviour before the probe reaches state A, but the discussion would equally apply to the continuation from A to a stable large amplitude state, B. Here there is no obstacle to progress along the probing path, and the whole of the displayed path is stable and can be explored satisfactorily. In particular, state A can be located, and the energy barrier given by the integral under the force-deflection curve is easily evaluated.

In the second probing characteristic, we display a local folding just after the appearance of two vertical tangencies (in a so-called cusp point). Such an event has recently been found and discussed in Section 3 (between A and B) in the post-buckling of a spherical shell when its volume is being rigidly controlled. After the main load $G$ is increased from (a) to (b) in Figure 12 through a 'cusp' there will be a small dynamic jump at constant probe deflection, which should restabilize on the same probing path. After this small jump, it might be possible to continue to state A as before, but such a jump would be dangerous and might trigger premature buckling.

In the third and fourth characteristics, an encounter with a symmetry-breaking pitchfork bifurcation is displayed which can be either supercritical or subcritical. This will, of course, only occur if the system contains a symmetry that can be broken. In both cases, the route to state A has been rendered unstable. One such pitchfork was in fact found (before A) in an earlier study of a model cylindrical shell [Thompson & Sieber, 2016]. In that study it was shown that the unstable probing path after the bifurcation can be stabilized and followed by the introduction of a suitably placed secondary probe tuned to zero force. Nevertheless, these symmetry-breaking bifurcations are potentially dangerous events, which might not be foreseen.

The really important event that is highlighted in this figure is shown in the fifth characteristic, where a trans-critical bifurcation is encountered on the probing path. This is the 'asymmetric' bifurcation studied experimentally and theoretically by Roorda [1965], and described for a propped cantilever in Thompson & Hunt [1973]: it has an imperfection sensitivity governed by a one-half power law. Immediately after the trans-critical bifurcation there will be two folds with vertical tangencies, which completely disrupt and break the probing process, as shown in the sixth diagram.

Figure 13 shows how, in the vicinity of a large cusp (for example), the vertical tangencies and trans-critical bifurcations can be typically encountered. This diagram also usefully shows how extra areas of the stable (red) surface can be explored by varying the controls simultaneously to follow curved paths in control space. This would for example allow an experimentalist to manoeuvre around a cusp.

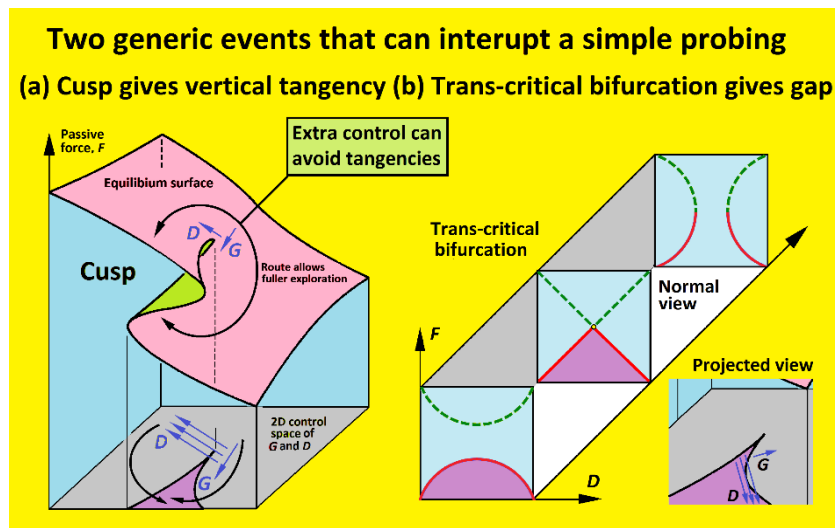

Fig 13 Surface features around a large cusp, showing the routes to vertical tangencies and a trans-critical bifurcation. Curved control routes show how more equilibria can be explored in synchronised control sweeps

We mentioned in the introduction that we might expect a lateral push on a shell to follow a sort of minimum energy path towards the unstable state, A. So how is it that we are now witnessing (in the spherical shell for example, in figures 5 and 6) vertical tangencies in which a path starts to return towards the displacement origin? Although it perhaps does not fully provide an explanation, it is useful to think about a ball being pushed by a frictionless plate over an energy landscape, following the concept introduced by Thompson & Sieber [2016]. The horizontal force maintains it original direction (perpendicular to the plate), and for equilibrium the ball must move to keep the plate tangential to the contours, and we see in figure 14 that a push up a curved valley can indeed give rise to a vertical tangency.

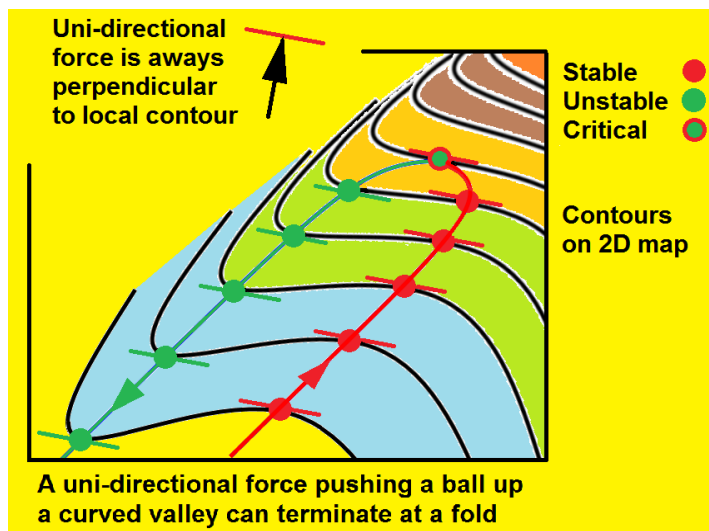

Fig 14. Sketch showing a ball pushed by a uni-directional force via a smooth plate. Setting off up a curved valley, the ball's path can turn back at what in our $(F, D)$ diagrams would be a vertical tangency.

## 8. General controllability through probing

While the experimental studies by Marthelot *et al* [2017] and Virot *et al* [2017] applied probing with a one-sided force (the poker in figure 10 pushes but cannot pull), the numerical results in figures 1, 3 and 5 extend the equilibrium surface for the probed shell to negative probe forces, achievable, for example, by gluing the poker to the shell (as proposed in the introduction). Gluing corresponds to a two-sided clamp, prescribing the displacement of the shell at the probed point. One might view the glued probe as an application of feedback control, since the glued probe can provide the force $F$ in real time depending on the displacement $D$. In feedback control terminology, the prescribed displacement $D$ is an additional input, while the force $F$ exerted at the clamp (measured, for example, with a stiff load cell) is an output.

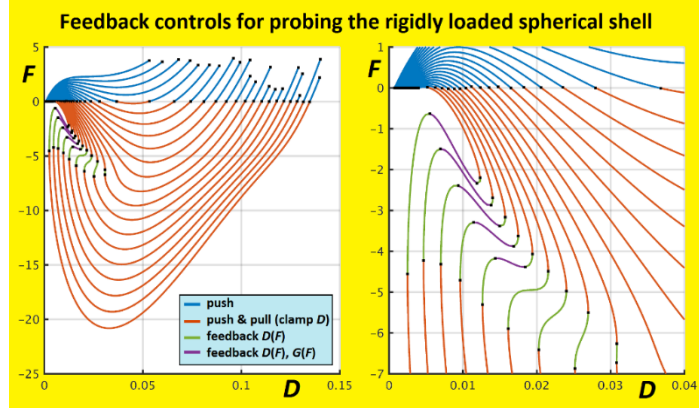

Fig 15. Front view of the probed shell equilibrium surface from numerical simulation, as also shown in figure 5. Shown are 21 probing responses for prescribed volume parameter, $G=dV/V_c$, evenly spaced from 0.18 to 0.98, with varying probe displacement $D$ and probe force $F$. On the right is a zoom-in close to $F$=0, $D$=0.

While the probing force landscape $F(D, G)$ depends on the specifics of the probe (for example, the location of the poker in figure 10) several its features are independent of the probe. As explained in the introduction, the curve defined by $F$=0 corresponds to the buckled states of the *unprobed* shell, since the probe has no effect by definition on the curve (no force is applied at the probe). This is a special case of the general principle that one may use feedback control in experiments to explore dynamically unstable phenomena: first introduce an artificial control input $u$ (in our case displacement $D$ of the probe), then vary $u$ and detect values $u_N$ for which the control is *non-invasive* (in our case, the measured force $F$ equals 0). The experiment for these values $u_N$ is in a state that corresponds to a natural state (for example, an equilibrium, in our case the buckled state) of the *uncontrolled* experiment. Thus, it may be worthwhile to introduce feedback control inputs into a pure test experiment (such as the buckling tests by Marthelot *et al* [2017] and Virot *et al* [2017]). The additional equipment and experimental effort required pays off in the ability to safely explore the experiment in states that are unstable, but important. In the case of buckling tests these are the unstable buckling states, which define thresholds to buckling of the shell. Figure 15 illustrates this principle. It shows a sequence of numerical simulations for probing tests that keep the volume prescribed ($G = \Delta V/\Delta V_C$ is evenly spaced between 0.18 and 0.98), while varying the probe displacement $D$ and measuring the force $F$. The underlying numerical results are of course identical to those in figure 5. Thus, all curves in figure 15 are on the probing force landscape of figure 5. If the probe is purely pushing only the part of the landscape surface covered by blue curves in Fig. 15 can be reached. Moreover, due to experimental disturbances one may have to stay away from the boundaries of the surface, which are precisely the buckling threshold states one is interested in. On the other hand, if the displacement $D$ is glued or clamped from both sides one can also access the part covered by red curves in figure 15 (in addition to the blue curves). As explained in Section 7, some obstacles in the landscape, such as cusps and folds can be avoided by the choosing suitable paths in the two-parameter $G$-$D$ plane, as illustrated in figure 13.

The interpretation of the probe as a type of feedback control gives guidance how one may overcome obstacles to exploring the entire probing force landscape by introducing additional feedback control inputs. Figure 15 illustrates the effect of improving feedback control. If one is able to adjust the displacement $D$ depending on the measured force in real time, then a simple linear feedback law of the form $D(F)=D_0+ g_D F$ will force the experiment to follow the correspondingly tilted line in the $D$-$F$ plane of figure 15. Consequently, for a suitably chosen control gain, $g_D$, we can access also the parts of the landscape covered by green curves. The boundary of controllability is then the curve in the landscape where the normal has a zero $D$ component. If one introduces a further control input by varying also the prescribed volume parameter $G$ depending on $F$, then a linear feedback law of the form $D(F) = D_0 + g_D F$, $G(F) = G_0 + g_G G$ makes every point on the landscape accessible with a suitable choice of control gains $g_D$ and $g_G$ except those isolated points where the normal has zero $D$ and $G$ components. This includes the part of the landscape covered by purple curves in figure 15. While for the particular shell parameters used in the simulations for figures 5 and 15 no further intersections of the probing force landscape with the force level $F$=0 occurred, this may be possible for other system parameters. For example, the local maxima of the $D$-$F$ curves for large $G$ (close to the critical volume), visible in the zoom on the right graph of figure 15 come close to $F$=0.

While the illustration in figure 15 focusses on feedback control through real-time variation of the control parameters $D$ and $G$, other additional feedback control inputs are possible. Thompson & Sieber [2016] suggested additional probing points. These additional probing points are feasible in a particularly straightforward way in experiments with simpler structures such as arches. Harvey & Virgin [2015] give an experimental demonstration of probing for shallow arches (though with only unilateral, one-sided clamping such that no unstable buckling states could be directly observed in the experiments).

In summary, the motivation for introducing feedback control into a test experiment is that it permits one to safely explore threshold states. The general principle is that one introduces feedback control inputs, which stabilize the experiment and then adjusts the inputs such that the actual control force at all inputs equals zero. This turns the test experiment into a nonlinear system of equations of dimension $n$ for $n$ control inputs. The process of adjusting the control inputs follows the same rules as one would use to solve any nonlinear system (for example, with a Newton iteration). Schilder *et al* [2015] give detailed instructions how one can modify standard numerical methods to cope with large disturbances as one might typically encounter in experiments. They also demonstrate feedback control-based bifurcation analysis in vibration experiments (other recent demonstrations and reviews on vibration experiments are listed by Barton [2017]).

## 9. Concluding remarks

Because of imperfection-sensitivity and the potential for catastrophic failure, the design of shell structures against buckling has traditionally relied heavily on empirical rules derived from many tests, often large-scale tests. Major efforts are currently underway to shift emphasis towards heavier reliance on computational methods. This shift is driven, in part, because many advanced shell structures are reinforced and stiffened in a multitude of ways making it increasingly difficult to rely on experimental data applicable to newly conceived design configurations. Consideration of the energy barrier against buckling is a new development in shell buckling. The barriers for unstiffened perfect spherical shells subject to pressure and for cylindrical shells under axial compression presented in figure 7 reveal the transition from a very weak barrier for shells loaded to within 30%, say, of the buckling load to a considerably larger barrier when the load is below about one half of the buckling load. The experimental results for the barrier obtained by displacement-controlled probing of the high quality spherical shell in figure 9 are in good agreement with the theory. The transition from a weak to a substantial barrier has clear implications for design against buckling. The general trends revealed here for probing perfect shells and for their energy barriers have been shown to carry over to imperfect spherical shells by Hutchinson and Thompson [2018].

   The present paper has addressed some of the complications that can arise in probing loaded shells with illustrations of specific examples where possible.  Probing a loaded shell to locate an unstable free buckling state and to measure the associated energy barrier can be a highly nonlinear process. The probe response can be interrupted by limit points and various types of bifurcation which, in turn, may inadvertently trigger dynamic buckling. Such events might possibly damage the test shell if steps (discussed here and in Thompson & Sieber [2016]) are not taken to control them. The recent probing experiments of Virot *et al* [2017] and Marthelot *et al* [2017] are particularly promising in that they have demonstrated that equilibrium buckled states can be located and the associated energy barriers can be quantified. The experiments were shown to exhibit reversibility and repeatability, but difficulty was sometimes experienced in getting precisely to the unstable state A. Additionally, it is notable that these experiments have been carried out on the two most imperfection-sensitive shell/loading systems known to exist.

   Imperfections determine the reduction of the buckling load below the prediction for the perfect shell while the energy barrier measures the robustness, or precariousness, of the shell in the loaded state to unanticipated loads and disturbances. Due to the localized nature of buckling of many shell structures, and especially for the unstiffened cylindrical and spherical shells, imperfections determining buckling will often be local. If the critical buckling state associated with the lowest applied load is to be located, the test probe must be applied in the vicinity of the imperfection. If, on the other hand, the test probe is applied in a region well away from the imperfection where the shell is relatively perfect, the test operator will measure a response that is more characteristic of the perfect shell than the real imperfect shell. While there is still much to learn about locating probes, a small number of examples illustrating such behaviour have been studied experimentally and theoretically in the papers referenced above. The lesson from this limited number of studies appears to be that, in addition to dealing with the potential complications that can arise in the probing response emphasized here, a probing protocol must also sample enough locations on the shell to ensure that the lowest buckling state is found.

……………… FF ……………….